\documentclass[prb,aps,twocolumn,amsmath,amssymb,floatfix,showpacs,
superscriptaddress]{revtex4}

\usepackage[dvips]{graphics}
\usepackage{color}
\definecolor{dred}{rgb}{0,0,0.6}

\begin{document}

\title{\textcolor{dred}{Magnetic response of non-interacting and 
interacting electrons in a M\"{o}bius strip}}

\author{Madhumita Saha}

\affiliation{Physics and Applied Mathematics Unit, Indian Statistical
Institute, 203 Barrackpore Trunk Road, Kolkata-700 108, India}

\author{Santanu K. Maiti}

\email{santanu.maiti@isical.ac.in}

\affiliation{Physics and Applied Mathematics Unit, Indian Statistical
Institute, 203 Barrackpore Trunk Road, Kolkata-700 108, India}

\begin{abstract}

We investigate characteristic features of both non-interacting and 
interacting electrons in a M\"{o}bius strip, the simplest possible 
one-sided topological system, in presence of an Aharonov-Bohm flux $\phi$.
Using Hartree-Fock mean field theory we determine energy eigenvalues 
for the interacting model, while for the non-interacting system an
analytical prescription is given. The interplay between longitudinal
and vertical motions of electrons along with on-site Hubbard interaction
yield several anomalous features of persistent current associated with
energy-flux characteristics. The variation of current with system size
and its temperature dependences are also critically examined. Current 
is highly sensitive to both these two factors, and we find that for a 
particular system size it decreases exponentially with temperature.
Our analysis can be helpful in investigating electronic transport through 
any non-trivial topological material.

\end{abstract}


\maketitle

\section{Introduction} 

The physics of topologically non-trivial materials can offer a new route
to design conventional electronic devices. NbSe$_3$ M\"{o}bius strip is 
one such possible geometry that was developed experimentally by Tanda
{\em et al.}~\cite{tanda} in $2002$ considering niobium and selenium 
compound. It is a one-sided topological system, unlike a regular cylinder, 
which is built by twisting a two-leg ladder and connecting its two ends. 
\begin{figure}[ht]
{\centering \resizebox*{4.5cm}{3cm}{\includegraphics{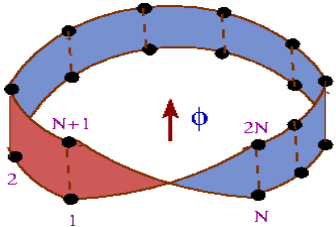}}\par}
\caption{(Color online). One-fold twisted M\"{o}bius strip threaded by an AB 
flux $\phi$ where the filled black circles correspond to the lattice sites.}
\label{f1}
\end{figure}
Several spectacular features are exhibited by a twisted 
M\"{o}bius geometry. One simplest and realistic example of such system can 
be the M\"{o}bius graphene strip. In $2009$ Guo {\em et al.}~\cite{gu} have 
shown that a
M\"{o}bius graphene strip with a zigzag edge behaves as a {\em topological 
insulator} with a gapped bulk and a robust metallic surface. Due to the 
significant potential applications, topological insulating materials have 
been under great focus both theoretically and experimentally, and the
M\"{o}bius graphene strip is the suitable candidate for it.
After successful fabrication of NbSe$_3$ inorganic conductor much attention 
has been given to explore electronic properties of different topological 
shape conductors~\cite{topo1,topo2,topo3,topo4,topo5}, expecting their 
strange contributions in designing nano-electronic devices. For a 
purposeful design, a clear
understanding of electronic behavior is highly important, and for isolated
conductors having single or multiple loops it can suitably be done by 
analyzing magnetic response in presence of Aharonov-Bohm (AB) flux $\phi$. 
Due to flux $\phi$, a Berry phase is introduced in moving electrons which
breaks time reversal symmetry and results a non-vanishing charge current. 
This is the so-called {\em persistent current}, an obvious demonstration of
AB effect, and was first proposed by B\"{u}ttiker {\em et al.}~\cite{butt1} 
during early $80$'s. Following this pioneering work, substantial theoretical 
and experimental works~\cite{gefen,cheu2,mont,bouz,giam,san1,san2,
madhu,ambe,schm1,schm2,bary,levy,jari,bir,chand,mailly1,mailly,blu} have 
been done along this line to understand different aspects of persistent 
current and other related issues in isolated conducting loops. 
In particular,
the physics of quantum rings has always been the subject of intense
research due to its potential applications in designing electronic, 
spintronic, optoelectronic and information processing devices. The 
innovative recent advances in experimental and theoretical physics of 
quantum rings are available in Ref.~\cite{fom}.

A similar kind of non-decaying circular current is also obtained in other 
context~\cite{cir1,cir2,cir3,cir4} where a ring-shaped conductor is connected 
with source and drain 
electrodes. Exploiting the effect of quantum interference among electronic 
waves passing through different branches of a conducting junction (viz, 
source-conductor-drain junction) one can establish a net current, in 
presence of a finite bias, which exhibits several interesting results. 
These features are not discussed here as they are beyond the scope of our 
present work, and hopefully we will reproduce them in our forthcoming work.

Now, the works involving flux-driven persistent current in isolated
systems are mostly confined to simple loop geometries like single-channel 
rings, multi-channel cylinders, graphene rings, nanotubes, array of rings 
to name a few~\cite{gefen,cheu2,mont,bouz,giam,san1,san2,madhu,ambe,
schm1,schm2,bary}. Whereas, very few works are available where twisted 
geometries have been taken into account. For instance, in $2003$ Cohen 
{\em et al.}~\cite{mob1} have studied the behavior of persistent current 
for a {\em non-interacting} M\"{o}bius strip and latter in $2009$ 
considering {\em spinless interacting electrons} Mori and Ota~\cite{mob2} 
have investigated electronic behavior in this particular geometry. 
In $2010$ Lassen {\em et al.}~\cite{nw1} have investigated 
finite-thickness 
effects considering different sized M\"{o}bius structures in presence of 
hydrostatic strain and explored several significant results. Latter in 
$2012$ Li and Ram-Mohan have done~\cite{nw2} a detailed study on a 
M\"{o}bius ring and revealed significant new ideas. In the same year, 
Fomin {\em et al.} have shown the delocalization-to-localization~\cite{nw3} 
transition taking an inhomogeneous M\"{o}bius ring which certainly 
highlights a great challenge in the current era of nanofabrication.
Though the studies involving electronic properties in different topological 
conductors have generated a wealth of literature knowledge, to the best of 
our knowledge, no one has reported the phenomenon of persistent current 
in presence of on-site Coulomb interaction which always gives non-trivial 
features and certainly it demands further study. 

In the present work we essentially focus on magnetic response of interacting
electrons in a M\"{o}bius strip where the interaction parameter is treated
within a Hartree-Fock (HF) mean field (MF) level~\cite{ee1,ee2,ee3,ee4,ee5}. 
The interplay between 
longitudinal and vertical motions of electrons along with on-site Hubbard
repulsion exhibits several anomalous features those are interesting and 
important too. Restricting electron motion along the vertical direction 
conventional $\phi_0$ ($=ch/e$, the elementary flux-quantum) periodicity 
of current can be changed to $\phi_0/2$, which was reported in the 
literature~\cite{mob1} considering non-interacting M\"{o}bius geometry. 
Apart from this, $\phi_0/2$ periodic
current can also be noticed depending on system size and filling factor 
even when the vertical motion is allowed. In addition we find that current
is highly sensitive to system size and temperature. Both these effects are
analyzed in detail. Though our main intention of the present work is to 
study magnetic response of interacting M\"{o}bius strip, for the sake of 
completeness here we also discuss characteristic properties of non-interacting 
electrons for which energy eigenvalues and persistent currents are evaluated
analytically. In absence of vertical hopping we can find closed analytical
form of net current for any arbitrary filling. Quite interestingly we see 
that for all odd number of electrons current gets a single expression, and 
similarly, for even number of electrons it gets another form.

Our work is organized as follows. In Sec. II we present the model and the
method for theoretical calculations. The results are presented in Sec. III,
and at the end we conclude in Sec. IV. 

\section{Model and Theoretical Formulation}

Figure~\ref{f1} displays a $2N$-site M\"{o}bius strip which is formed by
twisting a two-leg ladder, where each leg contains $N$ atomic sites, and
connecting its two ends. A magnetic flux $\phi$, measured in unit of $\phi_0$
($=ch/e$) is allowed to pass through the centre of the strip such that 
electrons move in a field-free region. To describe this model we use 
tight-binding framework and in presence of on-site Hubbard interaction it 
reads as,
\begin{eqnarray}
\textbf{H}_{M} & = & \sum_{\substack{j=1 \\ \sigma=\uparrow,\downarrow}}^{2N} 
\epsilon_{j,\sigma} c^{\dagger}_{j,\sigma} c_{j,\sigma} + 
t \sum_{\substack{j=1 \\ \sigma=\uparrow,\downarrow}}^{2N}
\left[e^{i \theta} c_{j,\sigma}^\dagger c_{j+1,\sigma} + h.c.\right]
\nonumber\\
& + & t_{\perp} \sum_{\substack{j=1 \\ \sigma=\uparrow,\downarrow}}^{2N} 
c_{j,\sigma}^\dagger c_{j+N,\sigma} + U \sum_{j=1}^{2N} 
c_{j,\uparrow}^\dagger c_{j,\uparrow} c_{j,\downarrow}^\dagger 
c_{j,\downarrow}
\label{eq1}
\end{eqnarray}
where the meanings of different symbols are explained as follows. 
$\epsilon_{j,\sigma}$ is the on-site energy of an electron at $j$th site
with spin $\sigma$ ($\uparrow,\downarrow$) and $c_{j,\sigma}^{\dagger}$
($c_{j,\sigma}$) represents the creation (annihilation) operator.
$t$ represents the nearest-neighbor hopping integral for the longitudinal
motion of electrons, while it is $t_{\perp}$ for the vertical motion.
$\theta$ ($=2\pi\phi/N\phi_0$) is the phase factor due to AB flux $\phi$
and $U$ gives the on-site Hubbard interaction strength. Here we impose the
boundary condition $j+2N=j$.

For $U=0$ the system becomes a non-interacting one, and under this 
situation all the features can be analyzed quite easily. Whereas for the
interacting case (viz, $U\ne 0$) it is very hard to find energy eigenvalues 
directly by diagonalizing the full many-body Hamiltonian (Eq.~\ref{eq1}),
in particular for large $N$ and higher number of up and down spin 
electrons~\cite{san1,ee4}.
Therefore, to find the energy eigenvalues in the present article we use 
Hartree-Fock mean field approximation which essentially decouples the
many-body Hamiltonian into two non-interacting ones associated with up
and down spin electrons~\cite{ee3,ee4,ee5}. The effective MF Hamiltonian 
gets the form:
\begin{eqnarray}
\textbf{H}_{M}^{MF} & = & \textbf{H}_{M,\uparrow} + \textbf{H}_{M,\downarrow}
- U \sum_{j=1}^{2N}\langle n_{j,\uparrow}\rangle\langle 
n_{j,\downarrow}\rangle 
\label{eq3}
\end{eqnarray}
where $\langle n_{j,\sigma} \rangle=\langle c_{j,\sigma}^{\dagger} 
c_{j,\sigma}\rangle$. The non-interacting Hamiltonians 
($\textbf{H}_{M,\uparrow}$ and $\textbf{H}_{M,\downarrow}$) are parameterized
with effective site energies, while the other parameters ($t$ and 
$t_{\perp}$) associated with electron hopping remain unchanged. The effective
on-site energies are $\epsilon_{j,\uparrow} + U \langle 
n_{j,\downarrow}\rangle$ and $\epsilon_{j,\downarrow} + U \langle 
n_{j,\uparrow}\rangle$, respectively, for up and down spin electrons.

From these decoupled non-interacting Hamiltonians we can easily determine
energy eigenvalues and evaluate net energy of the system at absolute zero
temperature ($T=0\,$K) from the relation
\begin{equation}
E_0(\phi) = \sum_{i=1}^{N_{\uparrow}} E_{M,\uparrow}^i +
\sum_{i=1}^{N_{\downarrow}} E_{M,\downarrow}^i - U \sum_{i=1}^{2N} 
\langle n_{i,\uparrow} \rangle \langle n_{i,\downarrow}\rangle 
\label{eq7}
\end{equation}
where $E_{M,\uparrow}^i$'s and $E_{M,\downarrow}^i$'s are the energy 
eigenvalues of the non-interacting Hamiltonians. $N_{\uparrow}$ and 
$N_{\downarrow}$ correspond to the number of up and down spin electrons, 
respectively, which fix the total number of electrons in the system
$N_e=N_{\uparrow}+N_{\downarrow}$. For finite temperature, this relation
(Eq.~\ref{eq7}) gets modified where the contributions from all energy
levels are taken into account with proper weight factor governed by the
Fermi-Dirac distribution function. In this case we have to specify chemical
potential $\mu$, instead of $N_e$.

Once $E_0(\phi)$ is determined, the persistent current is obtained from the
expression~\cite{gefen,mont,bouz}
\begin{equation}
I(\phi) =  -c \frac{\partial E_0(\phi)} {\partial \phi}.
\label{equ8}
\end{equation}
Thus taking the first order derivative of ground state energy
with respect to flux $\phi$ persistent current is determined, and it is the 
general expression~\cite{gefen} for evaluating persistent current in a system
whether it is characterized by fixed number of electrons $N_e$ or constant 
chemical potential $\mu$. 
At absolute zero temperature, $E_0(\phi)$ is determined by taking the sum 
of lowest $N_{\uparrow}$ and $N_{\downarrow}$ energy eigenvalues associated
with total number of electrons $N_e$ ($=N_{\uparrow} + N_{\downarrow}$) or
chemical potential $\mu$ for each value of $\phi$, as other energy levels 
are not occupied by electrons. While, for the case of non-zero temperature, 
finite occupation probabilities are obtained for all energy levels (they 
are different depending on the energy eigenvalues). Here we characterize the
system by constant $\mu$ (for a specific $\mu$, $N_{\uparrow}$ and 
$N_{\downarrow}$ are determined self-consistently), instead of $N_e$, and 
calculate the occupation probabilities of all the energy levels having 
energies $E_{M,\uparrow}^i$ and $E_{M,\downarrow}^i$. Then multiplying 
the occupation probability and associated energy eigenvalue of each level 
and taking the sum of this product over all energy levels we calculate 
$E_0(\phi)$. 

\section{Results and Discussion}

Below we present our results which include characteristic features of 
non-interacting and interacting electrons in a M\"{o}bius strip. Throughout 
the analysis we measure energy parameters in unit of electron-volt (eV) and
calculate current in unit of $et/h$, where $e$ and $h$ are the fundamental 
constants. We set $c=1$.

\subsection{Zero temperature limit}

Let us begin with non-interacting M\"{o}bius strip setting its temperature 
to zero. For non-interacting spinless case, the TB Hamiltonian reads as,
\begin{eqnarray}
\textbf{H}_M & = & \sum_{j=1}^{2N} \epsilon_{j} c^{\dagger}_j c_j 
+ t \sum_{j=1}^{2N} \left[e^{i \theta} c_j^ \dagger c_{j+1} + h.c. \right]
\nonumber\\
 & & + t_{\perp} \sum_{j}^{2N} c_{j}^\dagger c_{j+N}
\label{eq13}
\end{eqnarray}
where different terms carry identical meanings as discussed above. For a 
perfect M\"{o}bius strip $\epsilon_j$'s are identical and we can set them
to zero, for simplification. Under this situation the energy eigenvalues are
obtained from the relation:
\begin{equation}
E_n = 2 t \cos\left[\frac{\pi}{N}\left(n+\frac{2\phi}{\phi_0}\right)\right] 
+ t_{\perp}\cos[n\pi]
\label{eq14}
\end{equation}
where $n$ is restricted within the range $-N\leq n<N$. From this relation we
can calculate the current carried by $n$th eigenstate as
\begin{equation}
I_n = -\frac{\partial E_n(\phi)}{\partial \phi} = \frac{4\pi e t}{N h} 
\sin\left[\frac{\pi}{N}\left(n+\frac{2\phi}{\phi_0}\right)\right]
\label{eq155}
\end{equation}
and thus for $N_e$ electron system net current becomes 
$I(\phi)=\sum\limits_{n=1}^{N_e} I_n(\phi)$.

\begin{figure}[ht]
{\centering \resizebox*{6cm}{4cm}{\includegraphics{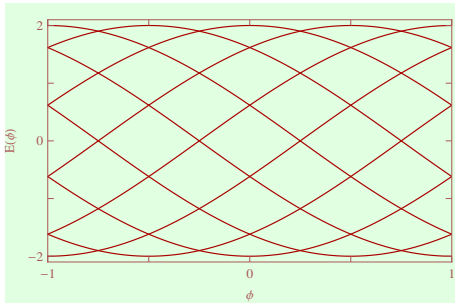}}\par}
\caption{(Color online). Energy-flux characteristics of a $10$-site 
non-interacting ($U=0$) M\"{o}bius strip with $t=-1$ and $t_{\perp}=0$.}
\label{f2}
\end{figure}

\begin{figure}[ht]
{\centering \resizebox*{6cm}{4cm}{\includegraphics{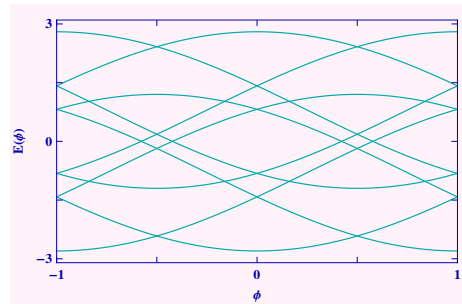}}\par}
\caption{(Color online). Energy spectrum for a $10$-site non-interacting
M\"{o}bius strip with $t=-1$ and $t_{\perp}=-0.8$.}
\label{f4}
\end{figure}

When $t_{\perp}=0$, the net current gets the form:
\begin{eqnarray}
I(\phi) & = & -2I_0 \frac{\sin\left[\frac{\pi}{2N}
\left(\frac{4\phi}{\phi_0}\right)\right]}{\sin\left[\frac{\pi}{2N}\right]}; 
~~ -0.25\leq \frac{\phi}{\phi_0}<0.25 \nonumber \\
 & & \hskip 4.6cm \mbox{for odd}~N_e 
\nonumber\\
 & = & -2I_0 \frac{\sin\left[\frac{\pi}{2N}
\left(\frac{4\phi}{\phi_0}-1\right)\right]}
{\sin\left[\frac{\pi}{2N}\right]}; 
~~ 0.0\leq \frac{\phi}{\phi_0} < 0.5 \nonumber\\
 & & \hskip 4.4cm \mbox{for even}~N_e
\label{eq144}
\end{eqnarray}
where $I_0=\frac{ev_f}{L}$, $v_f$ being the Fermi velocity determined 
at $k=k_f$ (Fermi wave vector) and $L=2Na$ ($a$ is the lattice spacing).
For non-zero $t_{\perp}$ we cannot find any such closed form of current for
a wide flux window, like Eq.~\ref{eq144}, for arbitrary electron filling.
It is apparent from Eq.~\ref{eq14} that the term involving 
$t_{\perp}$ does not contain any flux dependent term, so that its 
contribution on persistent current should be lifted after differentiating
the energy with respect to flux (Eq.~\ref{eq155}), and thus, one can also
expect the closed analytical form of current like Eq.~\ref{eq144} for the
situation when $t_{\perp}$ is finite. But this is not true, as the closed
analytical form of current is obtained only when the contributing energy 
levels (indexed by $n$) appear sequentially i.e., $n=0$ for $N_e=1$;
$n=0$, $-1$, $1$ for $N_e=3$, and similarly for even $N_e$. The appearance
of contributing energy levels within the above mentioned flux range for 
odd and even $N_e$ can be easily understood from the energy-flux spectrum
given below (see Fig.~\ref{f2}). This sequence of $n$ is not followed when 
$t_{\perp}$ is finite, as it changes the pattern of energy-flux 
levels (see Fig.~\ref{f4}) and that is reason behind the consideration 
of $t_{\perp}=0$ to get closed analytical form of persistent current
given in Eq.~\ref{eq144}.

\begin{figure}[ht]
{\centering \resizebox*{6cm}{7cm}{\includegraphics{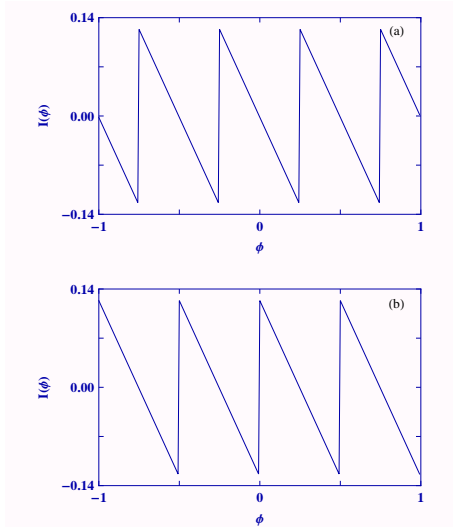}}\par}
\caption{(Color online). Persistent current as a function of flux $\phi$
in a $200$-site M\"{o}bius strip, where (a) and (b) correspond to $N_e=95$
and $96$, respectively. The other physical parameters are: $U=0$, $t=-1$
and $t_{\perp}=0$.}
\label{f3}
\end{figure}

Based on the above analytical expressions (Eqs.~\ref{eq14}-\ref{eq144})
we can easily characterize energy levels and current-flux spectra. 
In Fig.~\ref{f2} the full energy spectrum is shown for a $10$-site 
M\"{o}bius strip considering $t=-1$ and $t_{\perp}=0$. Multiple crossings
among different energy levels are obtained, yielding degeneracies, at
different values of flux like $\phi=0$, $\pm m\phi_0/4$, $\pm m\phi_0/2$
and $\pm m\phi_0$, where $m$ is an integer. All these energy levels
exhibit $\phi_0/2$ flux-quantum periodicity, unlike conventional $\phi_0$
periodicity obtained in a regular cylinder. The reason is that for 
$t_{\perp}=0$ an
electron which moves along the strip encloses a flux $2\phi_0$, instead of
$\phi_0$, when it comes back to its initial position as it encircles the
loop twice. This behavior gets reflected in current-flux characteristics.
To illustrate it in Fig.~\ref{f3} we present the variation of persistent
current in a $200$-site M\"{o}bius strip with $t=-1$ and $t_{\perp}=0$,
considering odd and even number of electrons.
Current exhibits saw-tooth like variation where sharp transitions at
different AB fluxes are associated with the crossing of energy levels. 
A clear signature of $\phi_0/2$ periodicity is observed from these 
$I$-$\phi$ spectra (Fig.~\ref{f3}).

The energy spectrum gets significantly modified with the inclusion 
of $t_{\perp}$. It is shown in Fig.~\ref{f4} where we plot distinct 
energy levels for a $10$-site M\"{o}bius strip considering $t=-1$ and 
$t_{\perp}=-0.8$. An overlap region appears across the energy band centre,
unlike a regular single-channel ring, which is responsible in producing 
anomalous kink-like structure in persistent current provided the Fermi
energy lies within this energy zone. The appearance of this overlap region 
can be explained from the energy expression given in Eq.~\ref{eq14}.
Depending on the value of energy level index $n$, Eq.~\ref{eq14} gets
splitted into two relations as 
\begin{equation}
E_n^1(\phi) = -t_{\perp} + 2t\cos\left[\frac{\pi}{N}\left(n + 
\frac{2\phi}{\phi_0}\right)\right]  
\label{eq15}
\end{equation}
and
\begin{equation}
E_n^2(\phi) = t_{\perp} + 2t\cos\left[\frac{\pi}{N}\left(n + 
\frac{2\phi}{\phi_0}\right)\right]
\label{eq16}
\end{equation}
These two expressions produce two energy sub-bands and their overlap 
is essentially controlled by $t_{\perp}$. For finite strength of 
$t_{\perp}$, electron does not acquire $2\phi_0$ flux to reach to its 
initial starting point, rather it encloses $\phi_0$ flux, which results 
energy levels $\phi_0$ periodic (see Fig.~\ref{f4}).

\begin{figure}[ht]
{\centering \resizebox*{6cm}{7cm}{\includegraphics{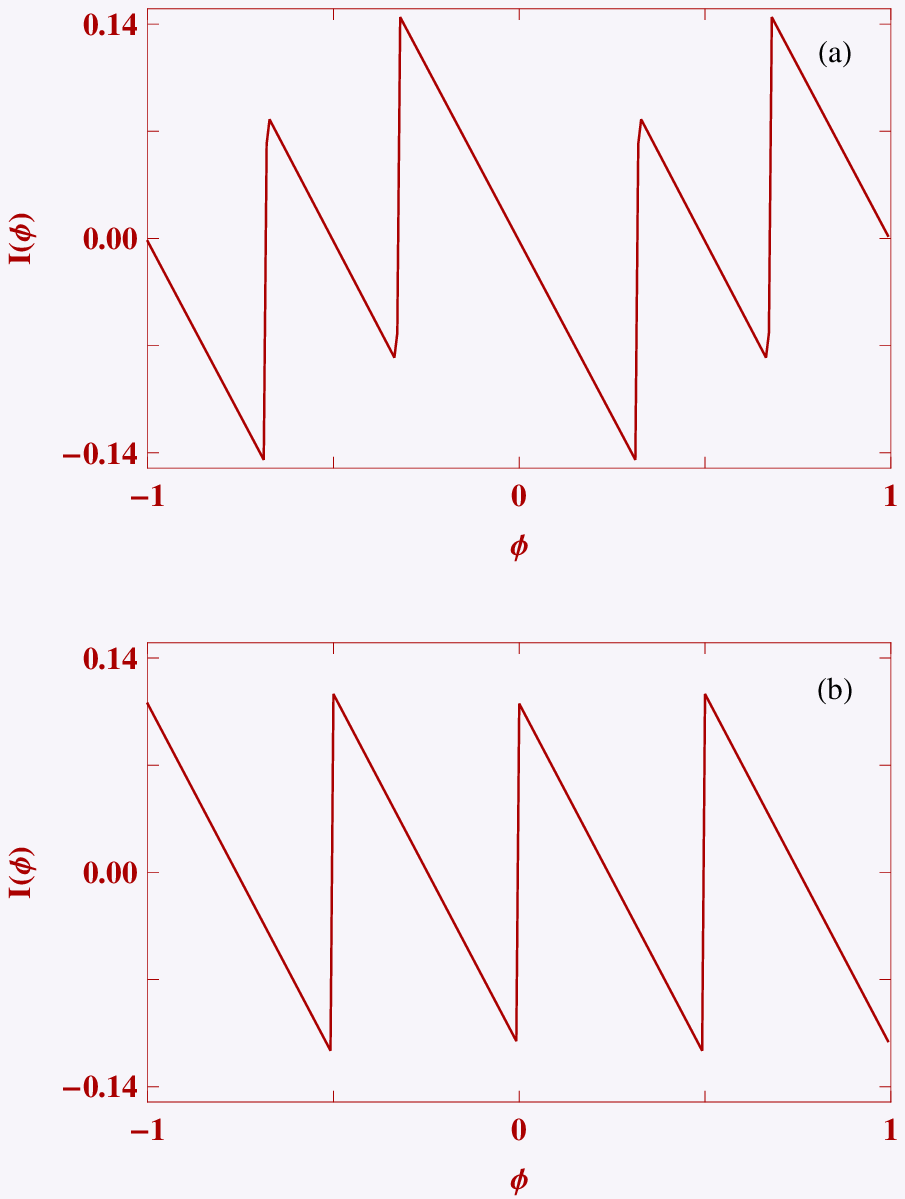}}\par}
\caption{(Color online). Current-flux characteristics of a $200$-site 
non-interacting ($U=0$) M\"{o}bius strip considering $t=-1$ and 
$t_{\perp}=-0.8$, where (a) and (b) correspond to $N_e=95$ and $96$,
respectively.}
\label{f5}
\end{figure}

This features enables us to characterize current-flux spectra given in 
Fig.~\ref{f5} where we present the variation of current as a function of
flux $\phi$ for a $200$-site M\"{o}bius strip considering $t=-1$ and 
$t_{\perp}=-0.8$, where (a) and (b) correspond to $N_e=95$ and $96$,
respectively. For odd $N_e$, a kink-like structure is observed across
$\phi=\pm 0.5$, while for even $N_e$ it is not separable from the other
parts due to increased kink height. In both these two fillings current
exhibits conventional $\phi_0$ periodicity, following $E$-$\phi$ curves
(Fig.~\ref{f4}).

\begin{figure}[ht]
{\centering \resizebox*{6cm}{4cm}{\includegraphics{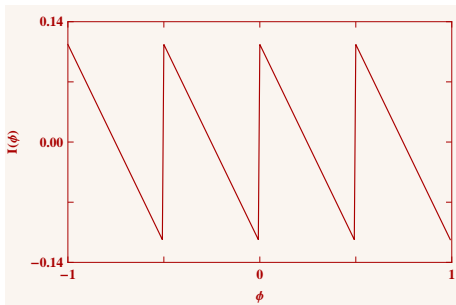}}\par}
\caption{(Color online). $I$-$\phi$ spectrum of a M\"{o}bius strip in the
half-filled band case. The parameters are: $N=100$, $U=0$, $t=-1$, 
$t_{\perp}=-0.8$. Current exhibits $\phi_0/2$ periodicity though $t_{\perp}$
is finite.}
\label{f6}
\end{figure}

Though $\phi_0$ periodic current is naturally expected for finite value
of $t_{\perp}$, but under a certain condition current yields half-flux
quantum ($\phi_0/2$) periodicity. It is the {\em half-filled} band case 
with {\em even} $N$. This is exactly what we present in Fig.~\ref{f6} where
current is computed for a $200$-site (i.e., $N=100$) M\"{o}bius strip in the
half-filled limit. Here it is important to note that for a regular cylinder 
(untwisted geometry), $\phi_0/2$ periodicity is also observed at 
half-filling but $N$ should be {\em odd}.

\begin{figure}[ht]
{\centering \resizebox*{6cm}{7cm}{\includegraphics{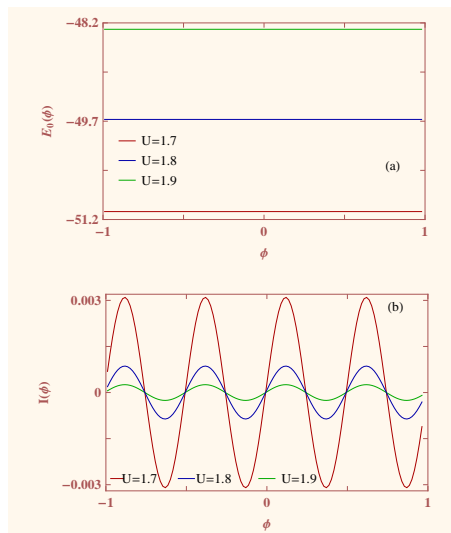}}\par}
\caption{(Color online). Dependence of ground state energy (upper panel) 
and corresponding current (lower panel) as a function of flux $\phi$ for a 
$60$-site interacting M\"{o}bius strip with $t=-1$ and $t_{\perp}=0$ for 
three different values of $U$. Here we choose 
$N_{\uparrow}=N_{\downarrow}=30$.}
\label{f7}
\end{figure}

The results analyzed so far are worked out for non-interacting M\"{o}bius
strips, and now we focus our attention on the behavior of interacting 
electrons. In Fig.~\ref{f7} we present the variation of ground state 
energy and corresponding persistent current as a function of flux $\phi$
for an interacting $60$-site M\"{o}bius strip in the half-filled band case 
($N_{\uparrow}=N_{\downarrow}=30$) for different values of $U$ considering 
$t=-1$ and $t_{\perp}=0$. It is found that with increasing the on-site 
Coulomb correlation strength $U$, ground state energy increases and its
slope also gets changed though it is not clear from the spectrum 
(Fig.~\ref{f7}(a)). This change in slope is nicely reflected in current-flux
characteristics (Fig.~\ref{f7}(b)), where we see that current varies 
periodically with $\phi$ providing $\phi_0/2$ periodicity and gets highly
suppressed with $U$. At half-filling all atomic sites are occupied by single 
electrons having a particular spin (up or down) which do not allow opposite 
spin electrons in the same site due to repulsive Coulomb interaction. Thus, 
the electronic hopping or more precisely electronic mobility gets suppressed 
which yields reduced persistent current. In the large $U$ limit we practically 
get zero current i.e., the system becomes a Mott insulator.

Even in presence of $t_{\perp}$ current amplitude gets decreased with $U$,
in the limit of half-filling, but the reduction of current is not as mush 
\begin{figure}[ht]
{\centering \resizebox*{6cm}{7cm}{\includegraphics{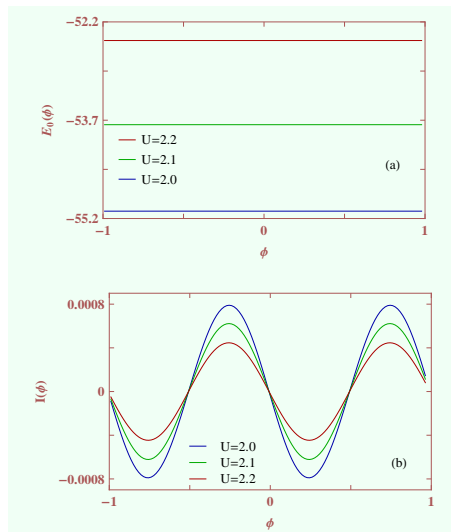}}\par}
\caption{(Color online). Ground state energy (upper panel) along with 
corresponding current (lower panel) as a function of $\phi$ for a $62$-site 
interacting M\"{o}bius strip with $t=-1$ and $t_{\perp}=-0.8$ for three 
different values of $U$. Here we choose $N_{\uparrow}=N_{\downarrow}=31$.}
\label{f8}
\end{figure}
as we get in the absence of $t_{\perp}$. The additional hopping (viz, 
$t_{\perp}$) is responsible for it. This behavior is clearly noticed from 
the results given in Fig.~\ref{f8} where we choose a $62$-site ($N=31$) 
interacting M\"{o}bius strip with $N_{\uparrow}=N_{\downarrow}=31$, $t=-1$ 
and $t_{\perp}=-0.8$. For this geometry (with odd $N$) the current exhibits 
usual one-flux quantum ($\phi_0$) periodicity as here we set a non-zero value 
of $t_{\perp}$. Whereas, an interacting half-filled M\"{o}bius strip with 
even $N$ exhibits unconventional half-flux quantum periodic current even 
though $t_{\perp}$ is finite (not shown here to save space), like what we 
get in the case of M\"{o}bius strip with non-interacting electrons 
(Fig.~\ref{f6}).

\begin{figure}[ht]
{\centering \resizebox*{6cm}{7cm}{\includegraphics{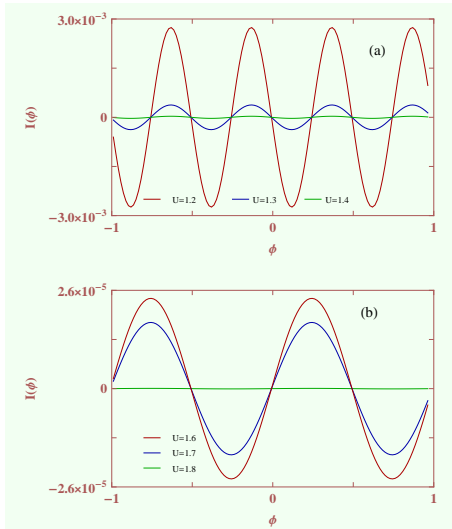}}\par}
\caption{(Color online). Current-flux characteristics for
a $102$-site interacting M\"{o}bius strip in the half-filled band case 
at three different values of $U$ where (a) $t=-1$, $t_{\perp}=0$ and
(b) $t=-1$, $t_{\perp}=-0.8$.}
\label{fnew}
\end{figure}

The results presented in Fig.~\ref{f7} and Fig.~\ref{f8} are
worked out for $60$- and $62$-site M\"{o}bius strips, respectively, which 
can be considered as ultra-small systems. To see the effect of interaction
among electrons in realistic M\"{o}bius rings which are significantly larger
in Fig.~\ref{fnew} we present the current-flux characteristics considering 
a $102$-site system in the half-filled band case.
Going through the spectra given in Figs.~\ref{f7}(b) and
\ref{f8}(b), it is clearly seen from Fig.~\ref{fnew} that the nature of
periodicity and the suppression of current with $U$ remain exactly same
for this $102$-site M\"{o}bius strip, and the nature will be exactly 
identical even for much higher system sizes which we confirm through our
detailed numerical calculation.
 
\begin{figure}[ht]
{\centering \resizebox*{6cm}{7cm}{\includegraphics{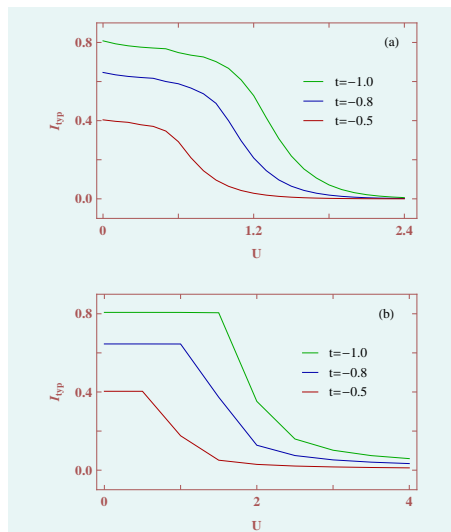}}\par}
\caption{(Color Online). $I_{\mbox{\tiny typ}}$-$U$ characteristics for
different values of $t$, setting $t_{\perp}=0$, for a $60$-site M\"{o}bius
strip, where (a) and (b) correspond to $N_{\uparrow}=N_{\downarrow}=30$
and $N_{\uparrow}=N_{\downarrow}=29$, respectively.}
\label{f9}
\end{figure}

\begin{figure}[ht]
{\centering \resizebox*{6cm}{7cm}{\includegraphics{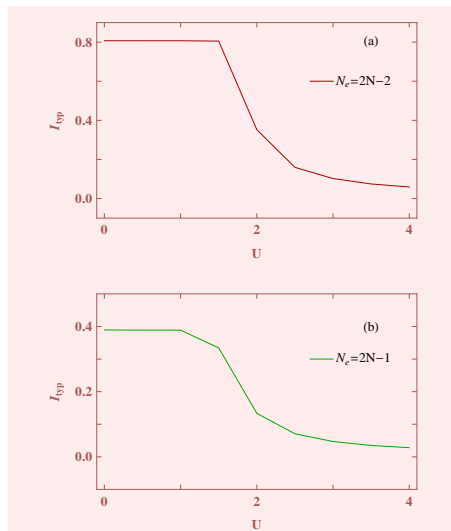}}\par}
\caption{(Color Online). Dependence of typical current amplitude 
$I_{\mbox{\tiny typ}}$ as a function of $U$ for four different band 
fillings considering a $60$-site ($N=30$) M\"{o}bius strip with $t=-1$
and $t_{\perp}=0$, where (a) and (b) correspond to even and odd $N_e$,
respectively. For even $N_e$ we set $N_{\uparrow}=N_{\downarrow}$, while
for odd $N_e$ we choose $N_{\uparrow}=N_{\downarrow}+1$.}
\label{f10}
\end{figure}

Now, in order to explain more clearly the interplay between on-site Hubbard 
interaction, hopping integrals, system size and filling factor we focus 
on the spectra given in Figs.~\ref{f9} and \ref{f10}, where the
variation of typical current amplitude is shown. Taking the absolute value
of maximum current within the range of one-flux quantum (viz, $0$ to 
$\phi_0$) we determine this typical current $I_{\mbox{\tiny typ}}$. 
Figure~\ref{f9} displays $I_{\mbox{\tiny typ}}$-$U$ characteristics for
different values of $t$ in the half-filled and less than half-filled band
cases for a $60$-site M\"{o}bius strip. The other hopping integral i.e.,
$t_{\perp}$ is fixed at zero. At half-filling current starts decreasing
when the e-e interaction is introduced, whereas for less than half-filled
case it ($I_{\mbox{\tiny typ}}$) remains almost constant for a specific
$U$-window and then decreases with $U$. These features are essentially 
controlled by two competing 
parameters $t$ and $U$. In the limit of half-filling each site of the 
system is occupied by an electron and thus movement of electrons is not
favorable due to repulsive interaction $U$ which results current reduction.
While, the presence of empty sites in less than half-filled system allows
electrons to hop from one site to other in the low $U$ limit where the 
hopping integral $t$ dominates over $U$ and makes the current almost $U$ 
independent. Beyond a critical $U$ repulsive interaction dominates and 
current starts decreasing. Eventually it reaches nearly to zero for large 
$U$. The rate of fall of current amplitude as well as the critical value 
of $U$ strongly depend on the filling factor, when the hopping integral 
remains constant, which is clearly seen from the spectra given Fig.~\ref{f10},
where (a) and (b) correspond to the even and odd $N_e$, respectively.
These features can be well understood from the above analysis. Both for
Figs.~\ref{f9} and \ref{f10} we compute the results setting $t_{\perp}=0$. 
Exactly similar features are also obtained even when $t_{\perp} \ne 0$ 
and that is why we do not present those results to save space.

\subsection{Finite temperature limit}

This sub-section discusses the effect of temperature on current-flux 
characteristics 
for both non-interacting and interacting M\"{o}bius geometries. 
\begin{figure}[ht]
{\centering \resizebox*{6cm}{7cm}{\includegraphics{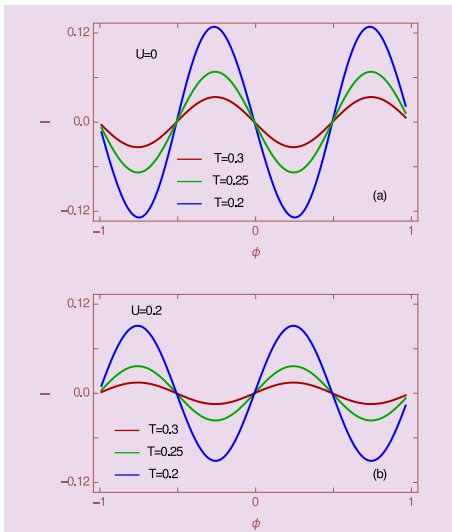}}\par}
\caption{(Color Online). Current-flux characteristics for a $60$-site 
M\"{o}bius strip at three typical temperatures, where (a) and (b) correspond
to $U=0$ and $0.2$, respectively. The other physical parameters are:
$t=-1$, $t_{\perp}=-0.6$ and $\mu=0.25$.}
\label{f11}
\end{figure}
In Fig.~\ref{f11} we present the variation of persistent current as a 
function of flux $\phi$ for three distinct temperatures considering
a $60$-site M\"{o}bius strip, where (a) and (b) correspond to the
non-interacting ($U=0$) and interacting ($U=0.2$) cases, respectively.
From the spectra it is observed that the current decreases with system
temperature. At finite temperatures, all energy levels contribute to 
current in certain percentage characterized by Fermi-Dirac distribution
function. With increasing the temperature occupation probabilities of higher
energy levels get increased and currents carried by successive energy 
levels in opposite directions are almost identical so that they mutually
cancel each other which results a smaller net current. Certainly much
lesser current is expected at higher temperatures. The reduction of current
due to repulsive Coulomb interaction (shown from the spectra given in
Fig.~\ref{f11}) remains same as discussed earlier.

\begin{figure}[ht]
{\centering \resizebox*{6cm}{4cm}{\includegraphics{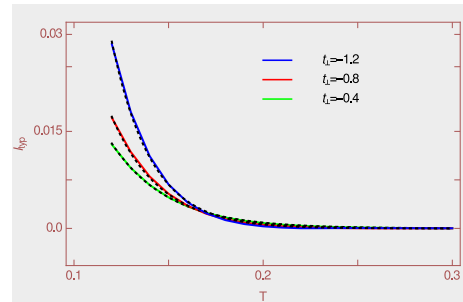}}\par}
\caption{(Color Online). Dependence of $I_{\mbox{\tiny typ}}$ with system
temperature $T$ for a spinless M\"{o}bius strip ($N=30$) at three different
values of $t_{\perp}$, setting $t$ and $\mu$ at $-1$ and $0.25$, 
respectively.}
\label{f12}
\end{figure}

Finally, to explore the asymptotic behavior of current with temperature 
we concentrate on the results presented in Fig.~\ref{f12}.
The variation of typical current $I_{\mbox{\tiny typ}}$ as a function of
temperature $T$ is given for a non-interacting spinless M\"{o}bius strip with
$60$ atomic sites considering $\mu=0.25$ and $t=-1$. Three different cases
are analyzed depending on $t_{\perp}$, where the colored dotted points
are computed from our theoretical prescription given in Sec. II. Using
these dots we find a functional relation between $I_{\mbox{\tiny typ}}$ and
temperature $T$ which looks like
$I_{\mbox{\tiny typ}}=a \exp(-bNT)$, where the constants $a$ and $b$ 
depend on $t_{\perp}$. For $t_{\perp}=-0.4$, $a=0.74$ and $b=1.12$, and
these values are $2$ and $1.32$ respectively for $t_{\perp}=-0.8$, and for
$t_{\perp}=-1.2$, these constant factors are $a=9.5$ and $b=1.61$. Plotting
this functional form we get the continuous curve, and we see that each curve, 
associated with $t_{\perp}$, matches extremely well with the dotted points. 
In this figure (Fig.~\ref{f12}) we present the results for a particular 
system size, but this exponential relation is absolutely general for any 
M\"{o}bius strip size which we confirm through our detailed numerical 
analysis. Only the factors $a$ and $b$, associated with $t_{\perp}$, get 
changed. In addition, it is important to note that even for interacting 
M\"{o}bius strip we find exactly identical functional relation of typical 
current with temperature $T$.
\vskip 0.25cm
\noindent
{\underline{\bf Accuracy of MF calculations:}} To make 
the present communication a self contained study, at the end, we would 
like to discuss about the accuracy of the mean-field calculations in our 
geometry.
\begin{figure}[ht]
{\centering \resizebox*{7cm}{5cm}{\includegraphics{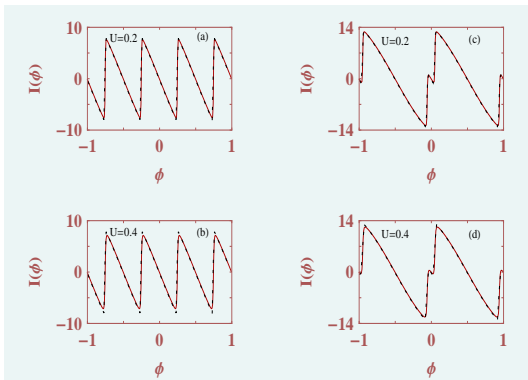}}\par}
\caption{(Color Online). Comparison between mean-field 
approach (black dotted line) and exact diagonalization method (red line)
is performed. 
Current-flux characteristics are computed for a $6$-site M\"{o}bius strip 
for two distinct values of $U$ in the limit of zero temperature, where 
the 1st column corresponds to $t_{\perp}=0$ and it is $-0.8$ for the 2nd 
column. Here we set $t=-1$.}
\label{accu}
\end{figure}
We make a comparative study by calculating persistent current
in two different ways. In one way we use Hartree-Fock mean field technique, 
and in the other way current is calculated by determining energy eigenvalues
through exact diagonalization of the full many-body Hamiltonian Eq.~\ref{eq1}. 
As the dimension of the Hamiltonian
matrix increases sharply with system size as well as up and down spin 
electrons, we restrict ourselves to a small system size due to our 
computational limitations in diagonalization. The results are presented 
in Fig.~\ref{accu}. The currents shown by black dotted lines are computed 
by exact numerical diagonalization 
method, while the MF results are shown by red curves. We see that MF results
match very well with the exact diagonalization technique. Here the currents 
are compared setting the system temperature at absolute zero. Similar 
agreement is also obtained for finite temperature, and thus, one can safely 
use HF mean-field approach to investigate magnetic response in our
twisted ring geometry.

\section{Closing Remarks}

In summary, we have investigated magnetic response of non-interacting and
interacting electrons in a one-fold twisted M\"{o}bius strip subjected to
an AB flux $\phi$. For the non-interacting system we have calculated
energy eigenvalues and the corresponding current completely analytically, and
under a typical case (viz, $t_{\perp}=0$) net current gets a closed form
within a specific flux window. For all odd $N_e$ it exhibits one 
particular relation, and similarly, for all even $N_e$ it follows another
relation. On the other hand, Hartree-Fock mean field theory has been utilized 
to study magnetic response of interacting electrons.

The essential findings are as follows. (i) Appearance of half-flux quantum 
($\phi_0/2$) periodicity when the vertical hopping between two ring-channels 
is restricted i.e., $t_{\perp}=0$. (ii) Even for non-zero value of 
$t_{\perp}$, $\phi_0/2$ periodic current can be observed if the system 
becomes half-filled and $N$ is even. (iii) Current is highly sensitive to 
the system temperature. It has been observed that, for a fixed system size, 
the typical current amplitude decreases sharply with increasing temperature 
$T$ following an exponential relation of the form 
$I_{\mbox{\tiny typ}}=a \exp(-b NT)$, irrespective of the e-e correlation 
strength.

In the present model we have ignored the effect of disorder. The interplay
between Hubbard interaction and disorder on persistent current has already
been discussed in several studies~\cite{mont,bouz,giam,ee4}, though mostly 
they are confined with simple loop geometries. Analogous behavior is also 
expected in M\"{o}bius
geometry, but a deeper insight into this problem is very essential for
further understanding. At the same time we have also ignored the effect of
electron-phonon interaction since it does not provide any significant change
in current in the said temperature regime.

Lastly we would like to state that all the features studied in this
article can be utilized to explore magnetic response in other non-trivial
topological systems.
 
\section{Acknowledgment}

MS is thankful to University Grants Commission, India for research 
fellowship.

\end{document}